\documentstyle[twocolumn,aps,psfig]{revtex}

\begin{document}

\newcommand{\PbPb}{\mbox{Pb+Pb}}
\newcommand{\SS}{\mbox{S+S}} 
\newcommand{\NN}{\mbox{N+N}}
\newcommand{\pp}{\mbox{p+p}}

\newcommand{\sqrtsNN}{\mbox{$\sqrt{s_{NN}}$}}

\newcommand{\Pb}{\mbox{$^{208}$Pb}}
\newcommand{\pmm}{\mbox{$p\!-\!\bar{p}$}}
\newcommand{\kmm}{\mbox{$K^+\!-\!K^-$}}
\newcommand{\pimm}{\mbox{$\pi^+\!-\!\pi^-$}}
\newcommand{\bmm}{\mbox{$B\!-\!\bar{B}$}}
\newcommand{\hminus}{\mbox{$h^-$}}
\newcommand{\kzeros}{\mbox{$K^0_S$}}
\newcommand{\lam}{\mbox{$\Lambda$}}
\newcommand{\lambar}{\mbox{$\bar{\Lambda}$}}
\newcommand{\lmm}{\mbox{$\lam\!-\!\lambar$}}
\newcommand{\sigzero}{\mbox{$\Sigma^0$}}
\newcommand{\sigzerobar}{\mbox{$\overline{\Sigma^0}$}}
\newcommand{\sigplus}{\mbox{$\Sigma^+$}}
\newcommand{\asigplus}{\mbox{${\bar{\Sigma}}^-$}}
\newcommand{\smm}{\mbox{$\sigplus\!-\!\asigplus$}}

\newcommand{\sigminus}{\mbox{$\Sigma^-$}}
\newcommand{\asigminus}{\mbox{${\bar{\Sigma}}^+$}}
\newcommand{\smmMinus}{\mbox{$\sigminus\!-\!\asigminus$}}

\newcommand{\piminus}{\mbox{$\pi^-$}}
\newcommand{\kplus}{\mbox{$K^+$}}
\newcommand{\kminus}{\mbox{$K^-$}}
\newcommand{\pbar}{\mbox{$\bar{p}$}}

\newcommand{\dedx}{\mbox{$\langle{dE/dx}\rangle$}}
\newcommand{\meandedx}{\mbox{$\langle dE/dx \rangle$}}
\newcommand{\meanpt}{\mbox{$\langle p_T \rangle$}}

\newcommand{\yprot}{\mbox{$y_{proton}$}}
\newcommand{\ycm}{\mbox{$y_{cm}$}}
\newcommand{\ybeam}{\mbox{$y_{beam}$}}

\newcommand{\pt}{\mbox{$p_T$}}
\newcommand{\ptot}{\mbox{$p_{tot}$}}
\newcommand{\mT}{\mbox{$m_T$}}
\newcommand{\ET}{\mbox{E$_T$}}

\newcommand{\meandy}{\mbox{$\langle \Delta{y} \rangle$}}
\newcommand{\lns}{\mbox{$\ln{(s)}$}}


\draft
\title{Baryon Stopping and Charged Particle Distributions in\\
 Central Pb+Pb Collisions at 158 GeV per Nucleon \\
}
\author{
H.~Appelsh\"{a}user$^{7,\#}$, 
J.~B\"{a}chler$^{5}$, S.J.~Bailey$^{16}$, L.S.~Barnby$^{3}$, J.~Bartke$^{6}$, 
R.A.~Barton$^{3}$, H.~Bia{\l}\-kowska$^{14}$, A.~Billmeier$^{10}$, 
C.O.~Blyth$^{3}$, R.~Bock$^{7}$
% ...
\\
(NA49 Collaboration)
}

\address {
$^{1}$Department of Physics, University of Athens, Athens, Greece\\
$^{2}$Lawrence Berkeley National Laboratory, University of California, Berkeley, USA\\
$^{3}$Birmingham University, Birmingham, England\\
$^{4}$KFKI Research Institute for Particle and Nuclear Physics, Budapest, Hungary\\
$^{5}$CERN, Geneva, Switzerland\\
$^{6}$Institute of Nuclear Physics, Cracow, Poland\\
$^{7}$Gesellschaft f\"{u}r Schwerionenforschung (GSI), Darmstadt, Germany\\
$^{8}$University of California at Davis, Davis, USA\\
$^{9}$Joint Institute for Nuclear Research, Dubna, Russia\\
$^{10}$Fachbereich Physik der Universit\"{a}t, Frankfurt, Germany\\
$^{11}$University of California at Los Angeles, Los Angeles, USA\\
$^{12}$Fachbereich Physik der Universit\"{a}t, Marburg, Germany\\
$^{13}$Max-Planck-Institut f\"{u}r Physik, Munich, Germany\\
$^{14}$Institute for Nuclear Studies, Warsaw, Poland\\
$^{15}$Institute for Experimental Physics, University of Warsaw, Warsaw, Poland\\
$^{16}$Nuclear Physics Laboratory, University of Washington, Seattle, WA, USA\\
$^{17}$Yale University, New Haven, CT, USA\\
$^{18}$Rudjer Boskovic Institute, Zagreb, Croatia\\
}

\date{\today}
\maketitle


\begin{abstract}

Net proton and negative hadron spectra for central \PbPb\ collisions
at 158 GeV per nucleon at the CERN SPS were measured and compared to
spectra from lighter systems. Net baryon distributions were derived
from those of net protons, utilizing model calculations of isospin
contributions as well as data and model calculations of strange baryon
distributions. Stopping (rapidity shift with respect to the beam) and
mean transverse momentum \meanpt\ of net baryons increase with system
size. The rapidity density of negative hadrons scales with the number
of participant nucleons for nuclear collisions, whereas their \meanpt\
is independent of system size. The \meanpt\ dependence upon particle
mass and system size is consistent with larger transverse flow
velocity at midrapidity for \PbPb\ compared to \SS\ central
collisions.

\end{abstract}

Lattice QCD predicts that strongly interacting matter at an energy
density greater than 1-2 GeV/fm$^3$ attains a deconfined and
approximately chirally restored state known as the quark-gluon plasma
(for an overview, see~\cite{qm96}). This state of matter existed in
the early universe, and it may influence the dynamics of rotating
neutron stars~\cite{glendenning}. The collision of nuclei at
ultrarelativistic energies offers the possibility in the laboratory of
creating strongly interacting matter at sufficiently high energy
density to form a quark gluon plasma~\cite{na49:caloPRL}. Hadronic
spectra from these reactions reflect the dynamics of the hot and dense
zone formed in the collision. The baryon density, established early in
the reaction, is an important factor governing the evolution of the
system~\cite{pkoch_kslee}. Comparison of model predictions with
measured rapidity and transverse momentum distributions and
correlation functions constrains the possible dynamical scenarios of
the reaction~\cite{hydro_theory}, such as those for longitudinal and
transverse flow~\cite{na49hbt}. In addition, the mechanism by which
the incoming nucleons lose momentum during the collision (baryon
stopping~\cite{BuszaGoldhaber}) is an important theoretical
problem~\cite{misc_theory,RQMD,VENUS}, and the measurement of baryon
stopping in heavy ion collisions provides essential data on this
question.

In this letter, we present measurements by the NA49 collaboration of
rapidity and transverse momentum distributions of participating
baryons and negative hadrons over a large fraction of phase space for
central
\PbPb\ collisions at 158 GeV per nucleon ($\sqrtsNN=17.2$
GeV). Hadronic spectra from \SS\ collisions at 200 GeV per
nucleon~\cite{na35:spectra,na35:new,na44:slopes} and
\NN\ (nucleon-nucleon) collisions at 200 and 
400 GeV~\cite{Gaz89,NNprotondata} serve as important references,
helping to identify effects that depart from those expected from the
linear superposition of many \NN\ collisions.

The NA49 apparatus is described in~\cite{na49:tpc}. A beam of
\Pb\ ions at 158 GeV per nucleon ($y_{lab}=5.8$) from the CERN SPS
struck a 224 mg/cm$^2$ thick natural Pb target. 
The 5\% most central collisions were
selected by measurement of the forward-going energy in the phase space
occupied by the projectile spectator nucleons. The reaction products
passed through a dipole magnetic field, and tracks from charged
particles used for this analysis were measured in two large Main Time
Projection Chambers (MTPCs) placed downstream of the magnets on either
side of the beam axis. Over 1000 tracks were measured in each
event. Identification of protons utilized the specific ionization
(\dedx) of the gas of the MTPC~\cite{dedx}. A relative
\dedx\ resolution of 6\%\ was achieved.

The evolution of the incoming baryons (baryon stopping) was studied
through measurement of the difference of the proton and antiproton
distributions (net protons, denoted \pmm) to eliminate the effect of
baryon-antibaryon pair production. Meson production was studied
through the yield of negative hadrons (\hminus), comprising primarily
\piminus, with an admixture of
\kminus\ and \pbar. $5\!\times\!10^4$ central events were 
used for the \pmm\ 
analysis and $7\!\times\!10^3$ central events were used for the 
\hminus\ analysis.

The net proton yield was determined by an identification technique
designed to minimize the systematic errors over wide
acceptance~\cite{Toy98}. For each phase space bin ($y$,\pt), the distribution
of \dedx\ from negatively charged particles was subtracted from that
of positively charged particles, resulting in a distribution that
contained a peak with positive amplitude (more positive than negative
tracks, principally due to the net proton yield), and a peak at higher
\dedx\ with negative amplitude (net pions). In order to extract the
particle yields, Gaussian functions were fitted to the two extrema in
each difference distribution. A correction to the net proton yield for
the difference between \kplus\ and \kminus\ yields, based upon
NA49 data~\cite{na49:kaons}, was 15\% at $\ycm\!<\!0$ and
was negligible for $\ycm\!>\!1$. The negative hadron yield was
determined from the yield of all tracks from negatively charged
particles excluding electrons, identified via \dedx.

Corrections for detector acceptance and track reconstruction
efficiency were calculated by embedding and reconstructing simulated
tracks in real events~\cite{Toy98}. The tracking efficiency was
greater than 95\% over most of the acceptance, falling below 80\% in
regions of high track density and near the edges of the
acceptance. Yield as a function of rapidity was determined only for
bins having acceptance to \pt=0, with the exception of the two
lowest rapidity bins for \hminus where extrapolation was based upon
the adjacent bin having full coverage. At high \pt, the acceptance was
limited to the region in which detector effects were well understood in
this analysis, extending at least to \pt=2.0 GeV/c for all rapidity bins.
Instrumental background due to secondary interactions with detector
material was estimated to be 5\% of the total measured yield, using a
GEANT-based Monte Carlo simulation and the VENUS event
generator~\cite{VENUS}, which reproduces transverse and forward energy
distributions in central \PbPb\
collisions~\cite{na49:caloPRL}. Excluding the decay corrections
discussed below, the systematic errors of all rapidity distributions
were less than 10\%.

The measured yields contain contributions from the products of weak
decays that were incorrectly reconstructed as primary vertex tracks.
The background correction to the \hminus\ yield due to the decay of
\kzeros\ was estimated using NA49 data~\cite{na49:kaons} to be 5\%\ at
midrapidity, decreasing strongly at higher rapidity. Background to the
\hminus\ yield due to \lam\ decay is less than 1\%. The background
correction to the \pmm\ yield is due to the decays of \lam, \lambar,
\sigplus, and \asigplus, and was assessed using a GEANT-based
simulation of the decay of \lam\ and \lambar\ and reconstruction of
their charged decay products. (\lam\ and \lambar\ should be understood
in this paper to include the contributions of \sigzero\ or
\sigzerobar\ as well as feeddown from weak decays, which are not
distinguishable from primary \lam\ or \lambar\ in this analysis.)  To
investigate the dependence of this correction on the phase space
distribution of the decaying particles, three different shapes of
rapidity distribution for \lmm\ were used, based upon (I) predictions
of the RQMD model~\cite{RQMD} and (II) the VENUS model~\cite{VENUS},
and (III) a preliminary NA49 measurement~\cite{na49:sqm97}. In each
case the \smm\ yield was assumed to be 30\%\ of the \lmm\
yield~\cite{Wro85} and to have the same rapidity distribution.
Strangeness is not conserved if both the hyperon yields in I or II and
the measured kaon distribution are used. The hyperon yields in these
cases were therefore scaled by the ratio of total charged kaon yields
in the data and the model, thereby imposing strangeness conservation.
In order to compare corrections due to \lmm\ distributions differing
only in shape but not total yield, the distribution in III was scaled
so that the ratio of \lam+\sigzero\ to charged kaon total yield agreed
with that from the models. The \pmm\ distribution corrected by other
\lmm\ distributions can be derived from the figure.

\begin{figure}
\centerline{\psfig{figure=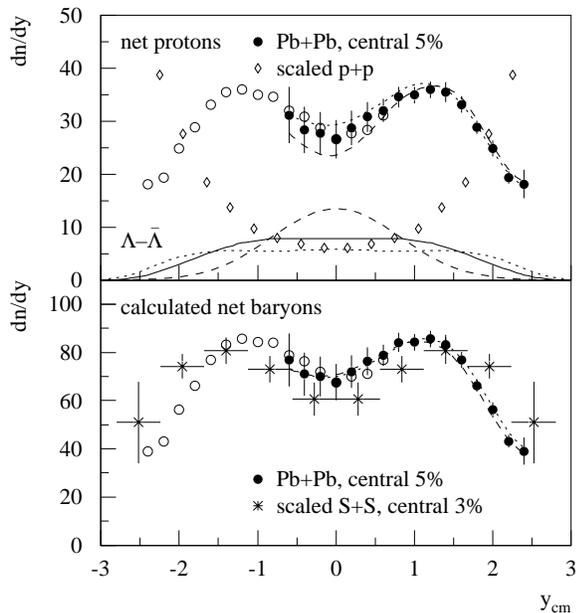}}
\caption{
Upper: normalized rapidity distributions of \pmm\ for \PbPb\
collisions incorporating correction I 
(open circles are data reflected about
\ycm=0, errors not shown). Lines 
show variation in data using corrections II (dotted) and III
(dashed). Also shown are corresponding \lmm\ rapidity distributions (I
solid, II dotted, III dashed), and 
the scaled proton distribution for \pp\ collisions.  
Lower: normalized rapidity
distributions of
\bmm\ from Eq.~\ref{eq:baryonnumber} for \PbPb\ incorporating
correction I, and scaled \bmm\ for \SS. Lines
correspond to corrections II (dotted) and III (dashed).
}
\label{fig:baryon}
\end{figure}

Fig.~\ref{fig:baryon}, upper panel, shows the event normalized net
proton yield as a function of rapidity for central \PbPb\ collisions,
incorporating the three \lmm\ corrections and corresponding \lmm\ rapidity
distributions. Also shown is the proton rapidity distribution for
\pp\ collisions at 400 GeV~\cite{NNprotondata}, which is
qualitatively different. The comparison highlights the importance of
multiple collisions to baryon stopping in nuclear collisions.

To determine the net baryon (\bmm)
rapidity distribution, the contribution of the remaining net baryons,
in addition to \pmm\ and
\lmm, must be estimated. Model calculations~\cite{RQMD,VENUS} indicate
that the rapidity distribution of net neutrons follows that of net
protons over most of phase space, with a $7\%$ larger yield. The
contributions of \smm\ and \smmMinus\ were accounted for by assuming
the same rapidity distributions as \lmm, and scaling the \lmm\ distribution 
by an empirical
factor $1.6\!\pm\!0.1$ derived from hadronic
reactions~\cite{Wro85}. Multistrange baryons contribute less than
2\%\ to \bmm~\cite{wa97}, and were not accounted for. The net baryon
yield \bmm\ was then calculated as
\begin{equation}
\bmm = (2.07\!\pm\!0.05)\cdot(\pmm)+(1.6\!\pm\!0.1)\cdot(\lmm).
\label{eq:baryonnumber}
\end{equation}

Fig.~\ref{fig:baryon}, lower panel, shows the \bmm\ rapidity
distribution for \PbPb. The data points and errors are for \lmm\
correction I. The variation in \bmm\ due to the use of different
\lmm\ corrections is smaller than that due to other sources of
systematic error. The \bmm\ yield is less sensitive than the
\pmm\ yield to the assumed \lmm\ distribution, because the \lmm\ distribution
is removed from the measured \pmm\ distribution by the decay
correction but added again in Eq.~\ref{eq:baryonnumber} to calculate
\bmm, with the two contributions approximately cancelling. Conclusions
on baryon stopping are therefore based upon the net baryon rather than
net proton rapidity distribution. Also shown in Fig.~\ref{fig:baryon},
lower panel, is the \bmm\ rapidity distribution for the 3\% most
central
\SS\ collisions at 200 GeV per nucleon~\cite{na35:new}, using a
coefficient of $2.0$ for
\pmm\ in Eq.~\ref{eq:baryonnumber} and with integral normalized to the
number of nucleon participants in \PbPb.  Within $|\ycm|\!<\!2.5$,
there are $352\pm12$ participant baryons for
\PbPb\ and $52\pm{3}$ for \SS\ central collisions~\cite{na35:new}.

The \bmm\ rapidity distribution is narrower for
\PbPb\ than for \SS\ collisions,
indicating increased baryon stopping for \PbPb\ collisions. (In the CM
frame, $\ybeam\!=\!2.9$ for \PbPb\ and $3.0$ for \SS.) The median
rapidity shift of the leading nucleon with respect to the beam in high
energy p+Pb collisions is 2-2.5 units~\cite{pA}. Due to the symmetry
of the nuclear reactions studied in this work, distributions of target
and projectile nucleon rapidity shifts in full phase space cannot be
determined separately, limiting the magnitude of a calculated rapidity
shift even in the case of large baryon stopping. The mean rapidity
shift relative to \ybeam\ for participant baryons within
$0\!<\!\ycm\!<\!2.5$ is $\meandy\!=\!-1.76\pm0.05$ for
\PbPb\ and $\meandy\!=\!-1.63\pm0.16$ for \SS\ collisions.

\begin{figure}
\centerline{\psfig{figure=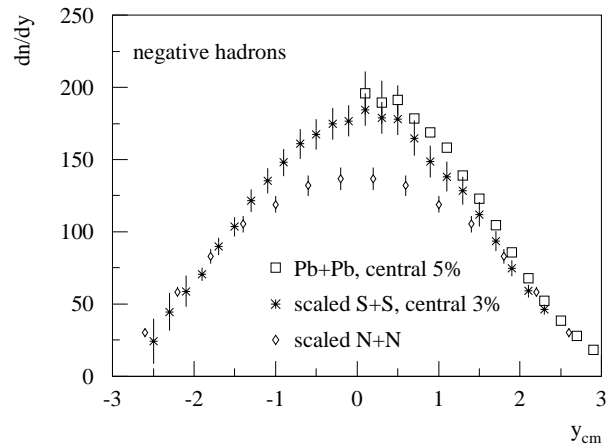}}
\caption{
Normalized rapidity distributions of \hminus\ for central \PbPb\
collisions, and scaled central \SS\ and isoscalar inelastic \NN\
collisions.
}
\label{fig:hminus}
\end{figure}

Fig.~\ref{fig:hminus} shows the event normalized rapidity distribution
(assuming pion mass) of \hminus\ in central \PbPb\ and
\SS~\cite{na35:new} collisions, and isoscalar inelastic \NN~\cite{Gaz89}
collisions. An independent analysis of \PbPb\ collisions using the
NA49 Vertex TPCs found a distribution consistent with these
data~\cite{Gue97}. The \SS\ and \NN\ distributions were scaled 
relative to that for
\PbPb\ by the ratio of the number of participant nucleons and a factor
0.96 to account for the energy dependence of average multiplicities
measured in hadronic collisions~\cite{lnS}. No correction for the
net isospin difference between \PbPb\ and the other systems was
applied. Under the assumption that the net isospin in the final state
of a \PbPb\ collision is carried entirely by mesons, the additional
scaling factor needed for comparison of the \SS\ and \NN\
distributions to that of \PbPb\ is 1.12.

The enhancement of meson yield at midrapidity for nuclear collisions
relative to \NN\ collisions has been noted
previously~\cite{na35:spectra}. Taking into account the net 
isospin difference, the agreement between the \PbPb\ and scaled \SS\
distributions is striking. This scaling of the yield of produced
particles with number of participants, first observed in p+A
collisions~\cite{pA}, is also observed in the collision of very heavy
ions. Extrapolation of the \hminus\ yield to full phase space was
performed by taking into account the asymmetry in the \hminus\ rapidity
distribution (assuming pion mass) due to the contribution of 
$K^-$~\cite{na49:kaons},
resulting in an estimated total \hminus\ yield of $695\pm{30}$
particles. The number of \hminus\ per participant nucleon pair, not
adjusted for isospin, is $4.0\pm0.2$ for \PbPb\ collisions, compared
to $3.6\pm0.2$ for \SS\ collisions and $3.22\pm0.06$ for isoscalar N+N
collisions~\cite{Gaz89}.

\begin{figure}
\centerline{\psfig{figure=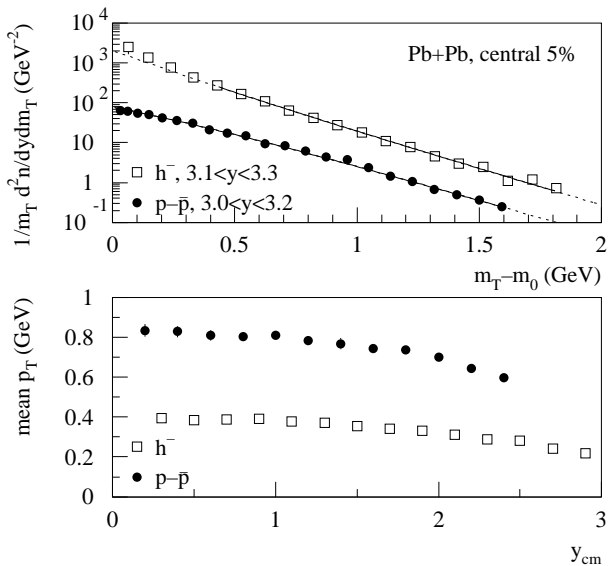}}
\caption{
Spectra for central \PbPb\ collisions. Upper: transverse mass spectra
near \ycm=0. Solid lines indicate fits discussed in text;
dotted lines are extrapolations of fit function. Lower: rapidity
dependence of \meanpt\ for \hminus\ and \pmm.
}
\label{fig:pt}
\end{figure}

Fig.~\ref{fig:pt}, upper panel, shows transverse mass spectra
($\mT=\sqrt{m^2+\pt^2}$) near midrapidity for \hminus\ and \pmm\ for
central \PbPb\ collisions. NA49 has previously reported the fit 
of an expanding hadronic source model~\cite{ChapmannScottoHeinz} 
to midrapidity \hminus\ and deuteron
\mT\ spectra and \hminus\ correlation functions~\cite{na49hbt}, 
giving a freezeout temperature of
$T\!=\!(120\pm12)$ MeV and transverse expansion velocity
$\beta_\perp\!=\!(0.55\pm0.12)$. Fig.~\ref{fig:pt}, upper panel,
shows the fit of this model with fixed $\beta_\perp\!=\!0.55$ to the
\hminus\ and \pmm\ spectra reported here. The resulting freezeout temperatures are
$T\!=\!(126\pm2)$ MeV for \hminus\ and $T\!=\!(118\pm5)$ MeV for \pmm.

Fig.~\ref{fig:pt}, lower panel, shows the rapidity dependence of
\meanpt\, calculated within $0\!<\!p_T\!<\!2.5$ GeV.
For \pmm\ for central \PbPb\ collisions, $\meanpt\!=\!(825\pm37)$ MeV at
$\ycm\!=\!0$ and $(600\pm17)$ MeV at $\ycm\!=\!2.4$, compared to the
rapidity-averaged value $\meanpt=(622\pm26)$ MeV for central \SS\
collisions~\cite{na35:spectra}. For \hminus\ for central
\PbPb\ collisions, $\meanpt\!=\!(385\pm18)$ MeV at $\ycm\!=\!0$,
compatible with $(377\pm4)$ MeV for central \SS\
collisions~\cite{na35:new}. (In considering \meanpt\ for \hminus\ at
high rapidity, note that the kinematic limit for pions for
\NN\ collisions at 158 GeV falls below $\pt\!=\!1$ GeV near beam rapidity).

A similar characterization of \pt\ distributions results from fitting
$(1/m_T)\cdot{dn/dm_T}$ with a function $A~\cdot~\exp{(-m_T/T)}$.  A
fit within $0\!<\!m_T\!-\!m_0\!<\!0.8$ GeV to the \pmm\ data near
midrapidity for central \PbPb\ collisions yielded $T\!=\!(308\pm15)$
(NA44 reported $(289\pm7)$ MeV for protons~\cite{na44:slopes}). NA35
found rapidity--averaged $T\!=\!(235\pm9)$ MeV for central \SS\
collisions~\cite{na35:new}. The increase in $T$ and \meanpt\ with
particle mass is consistent with a larger transverse radial flow
velocity at midrapidity in
\PbPb\ than \SS\ collisions~\cite{na49hbt,na44:slopes}.


In summary, baryon stopping for central collisions at
ultrarelativistic energies increases with system size. The integrated
yield and rapidity density of negative hadrons exhibit scaling with
the number of participant nucleons for nuclear collisions, and a small
enhancement with respect to \NN\ collisions. At midrapidity, a large
increase is observed in \meanpt\ for the stopped baryons in
\PbPb\ relative to \SS\ collisions, with no significant increase in
\meanpt\ for negative hadrons. The increase in \meanpt\ with particle mass 
is consistent with an increase in transverse radial flow velocity for
heavier colliding systems.


This work was supported by the Director, Office of Energy Research,
Division of Nuclear Physics of the Office of High Energy and Nuclear
Physics of the U.S.~Department of Energy under Contracts
DE-AC03-76SF00098 and DE-FG02-91ER40609, the U.S.~National Science
Foundation, the Bundesministerium f\"{u}r Bildung und Forschung,
Germany, the Alexander von Humboldt Foundation, the U.K.~Engineering
and Physical Sciences Research Council, the Polish State Committee for
Scientific Research (2 P03B 02615 and 2 P03B 9913), the EC Marie Curie
Foundation, and the Polish-German Foundation.

\vspace{.5cm}

$^{\dag}$deceased
$^{\$}$present address: Kent State Univ., Kent, OH, USA
$^{\#}$present address: Physikalisches Institut, Universitaet Heidelberg, Germany
$^{+}$present address: Max-Planck-Institut f\"{u}r Physik, Munich, Germany


\end{document}